
\def\MC{$\Omega (E, V) $}
\def\CPF{$Z(\beta,V)$}

\footline={\ifnum\pageno=1\firstfootline\else\otherfootline\fi}
\def\firstfootline{\rm\hss\folio\hss}
\def\otherfootline{\hfil}
\font\tenbf=cmbx10
\font\tenrm=cmr10
\font\tenit=cmti10
\font\elevenbf=cmbx10 scaled\magstep 1
 1
 1
\font\twelverm=cmr10 scaled\magstep 1

\overfullrule=0pt
\nopagenumbers
\line{\hfil}

\hsize=6.0truein
\vsize=8.5truein
\line{\hfil Brown HET-879, TA-483}
\vglue 1cm
\parindent=3pc
\baselineskip=10pt
\def\paris{\footnote{*}
{Invited talk at the Workshop on QCD Vacuum Structure, Paris
June 1992}}
\centerline{\tenbf IDEAL GAS OF STRINGS AND  QCD AT HADRONIC
SCALES\paris} \vglue 12pt \centerline{\tenrm CHUNG-I TAN} \baselineskip=13pt
\centerline{\tenit Physics Department, Brown University, Providence, RI 02912,
USA}

\vglue 0.8cm
\centerline{\tenrm ABSTRACT}
\vglue 0.3cm
{\rightskip=3pc
\leftskip=3pc
\tenrm\baselineskip=12pt
\noindent
By using lessons learned from modern string studies, we show how interesting
non-perturbative features of QCD can be learned from future heavy ion
collisions
even if the deconfinement density is not reached. \vskip20pt \vglue .8cm}

\line{\elevenbf 1.  Introduction\hfil} \vglue 0.4cm
\baselineskip=14pt
\twelverm
 It is the hope of many people that
experimental studies of ultra-relativistic nuclear collisions in the near
future
could reach  sufficiently  high energy densities so as to reveal the physics of
deconfined quarks and gluons. In this talk, I would like to suggest, based on
lessons learned from  modern string studies,$^{[1-3]}$ that
interesting non-perturbative features of QCD can already be learned even if the
deconfinement energy density, ${\cal E}_d$, is not reached.

It is well understood that the character of  QCD changes depending on  the
nature of available  probes. At short distances, the basic degrees of freedom
are quarks and gluons.  As one moves to larger distance scales, the QCD
coupling increases and one  enters the non-perturbative
regime.  Short of resulting to lattice Monte Carlo studies, the most promising
tool for a non-perturbative treatment of QCD  which builds in naturally
quark-gluon confinement remains the large-$N$ expansion. In this scheme,
although the vacuum of QCD at hadronic scales is complicated, model studies
suggest that  the effective degrees of freedom of QCD there can most profitably
be expressed in terms of ``extended objects". Indeed, low-lying hadron spectrum
suggests that they can be understood as ``string excitations".
  In high-energy soft
hadronic collisions$^{[4]}$ where the interactions are mostly peripheral, it is
possible to ``see" the dominant string excitations in terms of the exchanges
of high-lying Regge trajectories.

We would like to suggest that, in heavy ion collisions, it is possible to probe
the structure of string excitations inaccessible to other types of high energy
hadronic collisions. This can be accomplished even if heavy ion collision
experiments in the near future fail to  reach past the deconfinement scale
${\cal E}_d$.

 Assuming
thermo-equilibrium can be achieved in hadronic multi-particle production, one
expects that the energy density ${\cal E}$ can be parametrized monotonically in
terms of a temperature, $T$. At low temperatures,  ${\cal E}(T)$ can be given
effectively in terms of pion gas. At high temperature and after deconfinement,
one has Stefan-Boltzmann law, (appropriate for gluons and light quarks), which,
for $SU(2)$ flavor, leads to ${\cal E}\simeq 12 T^4$. Indeed, this expectation
has been substantiated by lattice calculations,$^{[5]}$ as depicted
below

\vskip210pt

{\rightskip=3pc
\leftskip=3pc
\tenrm\baselineskip=12pt
\noindent
Fig. 1. The energy density in finite temperature QCD with Wilson fermions,
taken from Ref. 5.  \vglue 0.8cm}

These ``numerical experiments"
suggest that the deconfinement density is of the order ${\cal E}_d\simeq 12
T_H^4$, where $T_H\sim 200 MeV$.
We would like to stress that these numerical experiments also indicate the
existence of another density scale, ${\cal E}_H\simeq c T_H^4$, where
$c=0(1)$.  We believe that, for   ${\cal E}_H< {\cal E}< {\cal
E}_d$ where the notion of temperature loses much of its usefulness,  the
effective degrees of freedom for QCD are  ``string-like".  Probing this
kinematical regime should reveal interesting non-perturbative features of
confined QCD at hadronic scales.$^{[6]}$
  \vglue 0.6cm
\line{\elevenbf 2.  Counting  Effective Degrees of Freedom at Hadronic
Scales\hfil} \vglue 0.4cm
 The basic degrees of freedom of a string theory are  string excitations.
Each excitation can be given a particle attribute, {\it i.e.}, a
mass $m$, with  its center of mass energy $p_0$ and spatial momentum vector
$\vec p$, $p_0^2=m^2+{\vec p}^2$. The  mass is due to internal string
oscillations; as such, it can take on increasing values, with a corresponding
increase in degeneracies.  For instance, the original dual model leads to an
operator expression for the mass squared,
$\alpha{'}m^2=-1+\sum_{i=1}^3\sum_{n=1}^{\infty}n{\hat N}_n^i$, where
$\alpha{'}$ is the ``Regge slope" and ${\hat N}_n^i$ is an ocillator number
operator taking on eigenvalues $0, 1, 2.\cdots.$ Well-known features of this
model include: (i) equal-spacing rule for the mass spectrum, and (ii)
exponentially increasing mass degeneracies, [{\it i.e.}, for $m^2=\alpha{'}N$,
the degeneracy
$d(m)=\sum_{N^i_n=0}^{\infty}\delta(N+1-\sum_{i=1}^3\sum_{n=1}^{\infty}nN_n^i)$
increases exponentially with $\sqrt N$.]

Formulating a consistent {\it effective  string theory} for QCD has been one of
the major challenges for string theorists.$^{[7]}$ Since we are still far from
accomplishing this task, any
insight into the problem, either theoretical or experimental, could prove
to be useful. A common feature of all string-like theories is the rapid
increase
of mass degeneracies. We shall assume that the desired effective QCD string
theory has an asymptotic  exponential mass degeneracy, which charaterizes the
growth of its effective degrees of freedom.  Alternatively, one finds that the
single-particle density behaves similarly at high energies
$$f(E,V)=V\sum_id(m_i)\int {d^3k\over h^3} \delta
(E-\sqrt{k^2+m_i})\propto  E^{-(\gamma+1)}e^{\beta_HE}.\eqno(1)$$
That is, we shall characterize a string theory minimally  by
 two critical exponents, $\gamma$ and $\beta_H$, (the latter is often referred
to as the inverse Hagedorn temperature.)

 It can be
demonstrated  that, for systems which can be charaterized by
Eq. (1), there always exists an energy density scale, ${\cal E}_H$, above
which the conventional canonical description becomes inapplicable.$^{[1-3]}$ A
microcanonical analysis can nevertheless be carried out which is valid both
above and below ${\cal E}_H$. On the other hand, under an ideal gas
approximation, the treatment is incapable of yielding the ``deconfinement
transition scale", ${\cal E}_d$,  which could come about only when interactions
are included. Since Monte Carlo experiments seem to indicate the existence of
a
sizable  density interval, $[{\cal E}_H, {\cal E}_d]$, in which  a large number
of string states are excited,   our subsequent analysis should be
meaningful there.

 \vglue 0.6cm
\line{\elevenbf 3.  Ideal Gas of Strings at High Energy Densities\hfil}
\vglue 0.4cm

For
statistical systems with a finite number of fundamental degrees of freedom, it
is well-known that a microcanonical and a canonical descriptions are
equivalent.  For strings, this  equivalence breaks down at high energy
densities.  However, it turns out that the canonical partition
function remains useful when considered as an analytic function of the
inverse temperature, $\beta\equiv 1/T$.

The fundamental quantity in a canonical approach is the partition function:
$Z(\beta,V)\equiv Tre^{-\beta  \hat H} =\sum_{\alpha} e^{-\beta E_{\alpha}},
$ where the sum is over all possibe multiparticle states of the system.
For a microcanonical approach, one works with
 a density function, which counts the number of microstates,   $\Omega (E,V)dE
\equiv \sum_{\alpha} {\delta (E - E_{\alpha})}\>\> dE$. Statistical mechanics
based on a microcanonical ensemble is more general, even though it is often
more
convenient to work with
 \CPF, {\it e.g.}, when interactions must be included.

 Representing the $\delta$-function by an integral along an
imaginary axis,
 we find that $\Omega(E,V)= \sum_{\alpha} \int_{-i\infty}^{+i\infty}
{d\beta\over{2\pi {i}}}
 \>\> e^{\beta \> (E-E_{\alpha})}$. Note that
the integral is in the form of an inverse Laplace transform over the
complex-$\beta$ plane. If one can deform the contour into a region where
interchanging the order of  sum and  integral is allowed, one obtains
$$\Omega(E,V)=\int_{\beta_0-i\infty}^{\beta_0+i\infty} {d\beta\over{2\pi {i}}}
 \>Z(\beta,V)\>\> e^{\beta E}. \eqno(2)$$ The allowed region is labelled by the
interception of the contour with the real axis, $\beta_0$.    One can then
recover \CPF\ from \MC\ via a Laplace transform,  $Z(\beta,V) =
\int_{0}^{\infty} dE \>\Omega(E,V)\>\> e^{-\beta E},$ which  provides an
alternative analytic definition for \CPF.

For conventional systems, Eq. (2) can often be approximated by a saddle point
 contribution at  $\hat\beta$.
For a closed system where $E$ is fixed, the usual notion of a temperature is
given by $\hat\beta^{-1}(\cal E)$, ${\cal E}\equiv E/V$, and is related to $E$
by the stationary condition: $E=-{\partial \log Z \over
{\partial \beta}}|_{\hat\beta}.$ We shall refer to this as the ``Boltzmann
temperature".

We have considered  elsewhere,  in a quantum statistical treatment,
analytic property of \CPF\ in $\beta$ for various string
theories.$^{[1]}$ We find that, generically, \CPF\ is
 analytic for $Re\>\beta>0$ except at isolated points. For each string
theory, because of the exponential growth in mass degeneracy, there is always
 an isolated rightmost singularity at $\beta=\beta_H$, {\it i.e.}, the
inverse Hagedorn temperature for that theory. There is a finite gap in their
real parts between $\beta_H$ and the next singularity to the left, and this gap
is theory-dependent but calculable.

To be more specific, for an ideal gas of
strings under the Maxwell-Boltzmann (MB) statistics, one has $Z(\beta, V)\simeq
e^{\tilde f(\beta, V)}$, where ${\tilde f(\beta, V)}$ is the inverse Laplace
transform of the single-particle density, $f(E,V)$. It can be shown that
$\tilde
f(\beta, V)$ is analytic for ${\rm Re}\> \beta$ sufficiently large and its
rightmost singularity is  at $\beta=\beta_H$. Given \CPF\  as an analytic
function of $\beta$, \MC\ can be recovered throught Eq. (2), with
$\beta_0>\beta_H$. That is, {\it  the totality of  physics of microcanonical
approach for free strings has been entirely encoded in the analyticity of
\CPF.}$^{[1]}$

   For strings, at low energies, Eq. (2) can be saturated by a
saddle point at $\hat\beta(E)$, lying to the right of $\beta_H$. However,
as the  energy density ${\cal E}$ is
raised, one reaches a point where either the saddle point moves to the left of
$\beta_H$, or it gets close to $\beta_H$ that the fluctuations about the saddle
point become large. When this occurs, the saddle point approximation to Eq. (2)
 break down, and it defines the  lower density scale ${\cal E}_H$
for the region of interest to heavy ion collisions spelled out earlier.

For ${\cal E}>{\cal E}_H$, whereas it is no longer meaningful to speak of a
Boltzmann temperature, the statistical mechanics of free strings is  still
given unambiguously by Eq. (2). One can in fact push the contour in (2)  to
the left of the singular point, $\beta_H$, by a finite distance $\eta$,
$\eta>0$. As one moves past this point, one picks up an additional contribution
involving the discontinuity across the cut. Denoting the discontinuity by
$\Delta{Z(\beta,V)}, \beta<\beta_H$,  the large-$E$ behavior of \MC\ is
dominated by the singularity at $\beta_H$
 $$ \Omega(E,V)=-\int_{\beta_H-\eta}^{\beta_H}
{d\beta\over{2\pi {i}}}  \Delta{Z(\beta,V)}e^{\beta
E}+0(e^{(\beta_H-\eta){E}}),
\hskip 30pt  \eta>0. \eqno(3)$$  Once $\Delta{Z(\beta,V)}$ is known, the
dominant behavior of \MC\ can be found. Therefore, the large-$E$ limit of a
free-string theory can best be approached by working first with the canonical
quantity, \CPF.$^{[1,2]}$  (Note that earlier related  works of Hagedorn and
co-workers$^{[5]}$  seem to have concentrated on the  approach to the region
${\cal E}\sim {\cal E}_H$ from below. Our analysis, on the other hand,
emphasizes on  the region above ${\cal E}_H$.)

\vglue 0.6cm
\line{\elevenbf 4.  Critical Exponent $\gamma$ and Classifications:  \hfil}
\vglue 0.4cm

We shall classify different types of  effective string theories
for confined QCD according by the  critical exponent $\gamma$ given in Eq.
(1). In standard string theories,  $2\gamma$ is equal to the
number of ``uncompactified" spatial dimensions. Therefore, the canonical value
is $\gamma=3/2$. However, since we are dealing with an effective string theory
for QCD, we could not rule out the possibility that this critical index can
take
on an anomalous value. Indeed, as we  suggest below, heavy ion collisions offer
the unique possibility of measuring this critical exponent. This  could in turn
provide  useful hint in our search for the realistic effective string theory
for
QCD.

It can be shown that, for $\gamma>0$,
$\tilde f(\beta)$ has a branch point at $\beta_H$ but it is bounded. For
$\gamma<0$, $\tilde f(\beta)$ has a divergent algebraic branch point. Finally,
when $\gamma=0$, $\tilde f(\beta)$ has a logarithmic branch point at $\beta_H$.
Model string theories have been constructed for both cases where $\gamma>0$ and
for $\gamma=0$. Instead of providing an exhaustive analysis, we shall assume
below that $\gamma\geq 0$.

For $\gamma>0$, $\tilde f(\beta)$  can be parametrized as $\tilde f(\beta)\sim
g(\beta)(\beta-\beta_H)^{\gamma}+\lambda(\beta)$, where
$\lambda(\beta)$ is regular at $\beta_H$. It follows that $Z(\beta)$ has the
same type of singularity at $\beta_H$, {\it i.e.}, $\Delta Z(\beta)\sim
(\beta-\beta_H)^{\gamma}$. For $\gamma=0$, one has $\tilde f(\beta)\sim
-c\log (\beta-\beta_H)+\lambda(\beta)$. We shall assume below that
$c=1$, as suggested  in the study of
fundamental strings.$^{[1-3]}$

The detailed statistical properties of a string gas turns out to be
sensitive to the value of $\gamma$. This can best be brought out by studying
the ``inclusive distributions" for our ideal gas of strings.    For instance,
the single-string distribution with a definite energy $\epsilon$ under the MB
statistics can be expressed as
$${\cal{D}}(\epsilon;E) = {1 \over {\Omega(E)}} f(\epsilon) [\Omega(E-\epsilon)
+
\delta(E-\epsilon)].\eqno(4)$$
The distribution ${\cal{D}}(\epsilon;E)$ is normalized so that, upon
integration over $\epsilon$, it yields the average number of strings, $\langle
N\rangle$,  in an ensemble with a total energy $E$, which can be more easily
measured experimentally.

\vskip
140pt  {\rightskip=3pc \leftskip=3pc \tenrm\baselineskip=12pt
\noindent
Fig. 2.  Schematic plots of $\epsilon{\cal D}(\epsilon, E)$ for a string
gas with (a): $\gamma\geq 3/2$,  and (b): $\gamma=0$. }
\vskip20pt

Exhaustive studies of this type can be found in Ref. 2,
where we make use of the analytic properties of $\tilde f(\beta)$ and
$Z(\beta)$
Here we shall mention several interesting features. For an ordinary gas of
particles, $\epsilon{\cal D}(\epsilon, E)$ as a function
of $\epsilon$ is typically $\epsilon^{\alpha}e^{-\beta\epsilon}$, {\it i.e.},
the usual Boltzmann distribution. In the case of string gas, for $\gamma\geq
3/2$, we find that the distribution is peaked at the two ends [Fig. (2a)] with
a
single energetic string soaking up most of the energy. This feature has been
referred  to in the past as the Frautschi-Carlitz  picture. In particular, one
finds that $\langle N\rangle$ becomes energy-independent at large $E$.   On the
other hand, for $\gamma=0$, we have a scale-invariant energy distribution, with
strings of all energies contributing equally to the total energy. One finds
that
$\langle N\rangle\sim \log E$, due to the $1/\epsilon$ tail of the distribution
at large energies. Experimental detection of this unique signature would be
most
interesting. (The situation is more complicated for $0<\gamma<3/2$. See Ref.
2.)

\vglue 0.6cm
\line{\elevenbf 5.  Comments:  \hfil}
\vglue 0.4cm
While the free string gas already exhibits novel properties, it is of
great interest to study the interacting gas. This could lead further insights
for describing the onset of deconfinement, a feature which is absent in an
ideal gas setting.  Finally, I have emphasized here on how lessons
learned from modern string studies could be used to study the physics of quarks
and gluons at confinement scales. It is of course also true that insights from
non-perturbative QCD studies, both theoretical and experimental, can also
provide new hints for works on fundamental strings. Increasing future
collaborative efforts in this direction could prove beneficial for both
communities.

\vglue 0.6cm
\line{\elevenbf 6.  References \hfil}\vglue 0.4cm

\item{1.}  Nivedita Deo, Sanjay Jain and Chung-I Tan,  {\it Phys.
Lett.} {\bf B 220}, 125 (1989).
\item{2.}  Nivedita Deo, Sanjay Jain and Chung-I Tan,  {\it Phys.
Rev.} {\bf D40}, 2626 (1989); {\it ibid.} {\bf D45}, 3641 (1992).
\item{3.}  J. Atick and E. Witten, {\it Nucl. Phys.} {\bf B310}, 291
(1988); R. Brandenberger and C. Vafa, {\it ibid.} {\bf B316}, 391
(1988); M. J. Bowick and S. Giddings, {\it ibid.} {\bf B325}, 631
(1989); P. Salomonson and B. S. Skagerstam, {\it ibid.} {\bf B268}, 349
(1986); N. Turok, {\it Physica} {\bf A158}, 516 (1989); Nivedita Deo, Sanjay
Jain and Chung-I Tan, ``The Ideal Gas of Strings", in Proceedings of the
International Colloquium on Modern Quantum Field Theory, Bombay, 1990 (World
Scientific) and references therein.
\item{4.} A. Capella, U. Sukhatme, C-I Tan, and Tran T. V., ``Dual Parton
Model", Orsay preprint, LPTHE 92-38, and references therein.
\item{5.}  H. Satz, {\it
Fortschr. Phys.} {\bf 33} (1985) {\bf 4}, 259-268, (in honor of  R. Hagedorn).
\item{6.} Atick and Witten, (Ref. 3), suggested that the deconfinement
transitio
in QCD should be  of the first order and would take place at a temperature
below
$T_H$, {\it i.e.}, in terms of the energy density, ${\cal E}_d<{\cal E}_H$. We
would like to suggest that  ${\cal E}_H<{\cal E}_d$, as indicated by the MC
data.  \item{7.} For  recent reviews: J. Polchinski, ``Strings and
QCD?",UTTG-16-92, and D. J. Gross, ``Some New/Old Approaches to QCD", LBL
33232, PUPT 1355.

\end